\documentclass[]{iopart}

\usepackage{epsfig}
\newcommand{\beq}{\begin{equation}}
\newcommand{\eeq}{\end{equation}}
\newcommand{\bes}{\begin{subequations}}
\newcommand{\ees}{\end{subequations}}
\newcommand{\bea}{\begin{eqnarray}}
\newcommand{\eea}{\end{eqnarray}}
\newcommand{\ba}{\begin{array}}
\newcommand{\ea}{\end{array}}
\newcommand{\beqn}{\begin{eqnarray*}}
\newcommand{\eeqn}{\end{eqnarray*}}

\newcommand{\f}[2]{\frac{#1}{#2}}

\newcommand{\hj}[1]{\vert \mbox{\boldmath{j}}_{#1}        \vert}
\newcommand{\hjvec}[1]{\hat{\mbox{\boldmath{j}}}_{#1}}
\newcommand{\Lvec} {\mbox{\boldmath{{L}}}}
\newcommand{\hjf}  {\vert \mbox{\boldmath{j}}_{\rm fin}   \vert}

\usepackage{graphicx}
\usepackage{graphicx,epsf, epsfig, amssymb}
\usepackage{bm}
\usepackage{longtable}

\def\be{\begin{equation}}
\def\ee{\end{equation}}
\def\beq{\begin{eqnarray}}
\def\eeq{\end{eqnarray}}
\def\f{\frac}

\def\gsim{\mathrel{
\rlap{\raise 0.511ex \hbox{$>$}}{\lower 0.511ex
\hbox{$\sim$}}}}

\def\lsim{\mathrel{
\rlap{\raise 0.511ex \hbox{$<$}}{\lower 0.511ex
\hbox{$\sim$}}}}

\begin{document}

\title[Massive Black Hole Binary Inspirals] 
{Massive Black Hole Binary Inspirals: Results from the LISA Parameter
Estimation Taskforce}

\author{
  K.~G.~Arun$^{1,2}$,
  Stas Babak$^3$,
  Emanuele Berti$^{4,5}$,
  Neil Cornish$^6$,
  Curt Cutler$^{4,7}$, 
  Jonathan Gair$^8$,
  Scott A.~Hughes$^9$,
  Bala R.~Iyer$^{10}$,
  Ryan N.~Lang$^9$,
  Ilya Mandel$^{11}$,
  Edward K.~Porter$^{3,12}$,
  Bangalore S.~Sathyaprakash$^{13}$,
  Siddhartha Sinha$^{10,14}$,
  Alicia M.~Sintes$^{3,15}$,
  Miquel Trias$^{15}$,
  Chris Van Den Broeck$^{13}$,
  Marta Volonteri$^{16}$
}

\address{$^1$ LAL, Univ. Paris-Sud, IN2P3/CNRS, Orsay, France}

\address{$^2$ Institut d'Astrophysique de Paris, UMR 7095-CNRS, Universit{\'e}
  Pierre et Marie Curie, 98$^{\rm bis}$ boulevard Arago, 75014 Paris,
  France}

\address{$^3$ Max-Planck-Institut f\"ur Gravitationsphysik (Albert-Einstein-Institut), Am M\"uhlenberg 1, D-14476 Golm bei Potsdam, Germany}

\address{$^4$ Jet Propulsion Laboratory, California Institute of Technology,
  Pasadena, CA 91109, USA}

\address{$^5$~Dept. of Physics and Astronomy, The University of
  Mississippi, University, MS 38677-1848, USA}

\address{$^6$ Dept. of Physics, Montana State Univ., Bozeman, MT 59717, USA}

\address{$^7$ Theoretical Astrophysics, California Inst.\ of Technology, Pasadena, CA 91125}

\address{$^8$ Institute of Astronomy, University of Cambridge, Cambridge, CB30HA, UK}

\address{$^9$ Dept. of Physics and Kavli Institute for Astrophysics and
  Space Research, MIT, Cambridge, MA 02139}

\address{$^{10}$ Raman Research Institute, Bangalore, 560 080, India}

\address{$^{11}$ Dept. of Physics and Astronomy, Northwestern Univ.,
  Evanston, IL, USA}

\address{$^{12}$ APC (AstroParticules et Cosmologie), 10, rue Alice Domon et
  L\'eonie Duquet, 75205 Paris Cedex 13, France}

\address{$^{13}$ School of Physics and Astronomy, Cardiff University, 5, The
  Parade, Cardiff, UK, CF24 3YB}

\address{$^{14}$ Dept. of Physics, Indian Institute of Science,
  Bangalore, 560 012, India}

\address{$^{15}$ Departament de F\'isica, Universitat de les Illes Balears,
  Cra. Valldemossa Km. 7.5, E-07122 Palma de Mallorca, Spain}

\address{$^{16}$ Dept. of Astronomy, University of Michigan, Ann Arbor, MI
  48109}

\begin{abstract}
  The LISA Parameter Estimation (LISAPE) Taskforce was formed in September
  2007 to provide the LISA Project with vetted codes, source distribution
  models, and results related to parameter estimation. The Taskforce's goal is
  to be able to quickly calculate the impact of any
  mission design changes on LISA's science capabilities, 
  based on reasonable estimates of the distribution of
  astrophysical sources in the universe. This paper describes our Taskforce's
  work on massive black-hole binaries (MBHBs). Given present uncertainties in
  the formation history of MBHBs, we adopt four different population models,
  based on (i) whether the initial black-hole seeds are small or large, and
  (ii) whether accretion is efficient or inefficient at spinning up the
  holes. We compare four largely independent codes for calculating LISA's
  parameter-estimation capabilities. All codes are based on the Fisher-matrix
  approximation, but in the past they used somewhat different signal models,
  source parametrizations and noise curves. We show that once these
  differences are removed, the four codes give results in extremely close
  agreement with each other. Using a code that includes both spin precession
  and higher harmonics in the gravitational-wave signal, we carry out Monte
  Carlo simulations and determine the number of events that can be detected
  and accurately localized in our four population models.
\end{abstract}




\section{Introduction}

These proceedings report results on merging massive black-hole binaries
(MBHBs) obtained by the LISA Parameter Estimation (LISAPE) Taskforce.  The
LISAPE Taskforce was established at the September, 2007, LISA International
Science Team (LIST) meeting at ESTEC, under the auspices of the LIST's Working
Group 1b (Data Analysis).  The LISAPE Taskforce was charged with developing a
set of vetted tools for quickly estimating LISA's science reach for various
mission configurations and gravitational-wave (GW) sources.

The initial impetus for creating our Taskforce arose because several research
groups had independently written codes to calculate LISA's capabilities to
extract the parameters of MBHBs
\cite{Cutler:1997ta,Moore:1999zw,Sintes:1999cg,Hellings:2002si,Vecchio:2003tn,Berti:2004bd,Berti:2005qd,Lang:1900bz,Arun:2007qv,Arun:2007hu,Trias:2007fp,Lang:2007ge,Porter:2008kn,Trias:2008pu},
but these groups' published results appeared on their face to be
discrepant. However, the various groups also used somewhat different
approximations in their signal models, as well as somewhat different
assumptions for the LISA noise curve (unfortunately, there has never been an
``official'' LISA noise curve). Therefore it was unclear whether the differing
results simply reflected these different assumptions, or whether they were due
to bugs in one or more codes.

The Taskforce's first goal was to resolve that question. As described in
Sec.~\ref{comparison}, it turned out that there were no bugs: the differing
results were primarily due to using different noise curves, and especially
different low-frequency cut-offs. The second goal was to arrive at vetted
parameter-estimation codes that the LISA Project could safely use in its work,
e.g., in helping set LISA's sensitivity requirements, as set forth in the
evolving LISA Science Requirements Document~\cite{ScRD}.

In addition, the Taskforce wanted to develop a set of models for the
distribution of MBHB events in the universe: event rates as a function of the
black-hole masses, spins, and redshift, based on well-motivated assumptions
regarding the birth and growth history of massive black holes in the universe.
The goal was to generate realistic source distributions, which we could
``feed'' to our vetted parameter-estimation codes in Monte Carlo fashion,
arriving at realistic ensembles of LISA observations and associated
parameter-estimation accuracies.  Our hope is that in the future other LISA
researchers will use these same MBHB ensembles when evaluating LISA science
performance, so that different researchers will be comparing ``apples to
apples''.

Of course, today our ignorance concerning MBHB birth and growth is rather
humbling \cite{Berti:2006ew,Sesana:2007sh,Volonteri:2007fz}.  It therefore
behooves us to consider a range of plausible assumptions. In the end, the
Taskforce settled on four representative source distributions, constructed
from four merger tree models.  The four merger-tree models arise from two
choices for astrophysical inputs: (1) whether the masses of the inital
``seed'' black holes are small or large, and (2) whether accretion is
efficient or inefficient at spinning up the massive black holes.  Many more
models could of course be developed, but the Taskforce felt that these four do
very broadly sample the range of possibilities, and so give useful insight
into how much LISA's science reach depends on the distribution of MBHBs in the
universe.

The organization and procedures of the LISAPE taskforce were modelled on those
of the quite successful Mock LISA Data Challenge (MLDC) Taskforce (see,
e.g.,~\cite{Babak:2008sn}). Membership is open, we have group telecons roughly
every two weeks, and the telecon minutes and other Taskforce documents are
posted on our wiki:

\vspace{.2cm}
http://www.tapir.caltech.edu/dokuwiki/lisape:home
\vspace{.2cm}

It turned out that the creation of our Taskforce was quite timely, since by
May, 2008, we were in a position to contribute significantly to a ``descope
exercise'' that the LISA project carried out in that month at NASA's
behest. The results of that exercise are summarized in R. Stebbins's
contribution to this volume.

The organization of this paper is as follows.  In Sec.~\ref{merger} we
describe the four merger-tree models that we use, and the underlying
astrophysical assumptions.  In Sec.~\ref{comparison} we describe the
inter-code checks that we did to validate our parameter-estimation codes.  All
these codes calculate expected parameter uncertainties via the Fisher-matrix
formalism~\cite{Vallisneri:2007ev}.  Such results are accurate only up to
corrections of order $1/\rho$, where $\rho$ is the signal-to-noise ratio
(SNR). However they are much faster to implement than more accurate Markov
Chain Monte Carlo methods, and so more suitable for exploration of a large
parameter space. For our results, presented in Sec.~\ref{results}, we used the
parameter-estimation code developed by the Montana/MIT group, since their
model waveforms already incorporated both higher harmonics and spin-orbital
precession effects.  In Sec.~\ref{conclusions} we summarize our conclusions
and discuss briefly the planned future work of the LISAPE Taskforce.

\section{Massive black-hole merger trees\label{merger}}

The cosmological evolution of massive black holes can be determined by
following the merger history of dark-matter halos and of the associated black
holes by cosmological Monte Carlo realizations of the merger hierarchy from
early times until the present in a $\Lambda$CDM cosmology
($H_0=70$~km~s$^{-1}$~Mpc$^{-1}$, $\Omega_{\rm M}=0.3$,
$\Omega_{\Lambda}=0.7$).
The simulations provide the masses, redshifts and spins of merging black
holes.

Two important sources of uncertainty in merger-tree models of black-hole
formation are (i) the formation mechanism and mass of the first ``seed'' black
holes, and (ii) the details of how accretion causes black holes to grow in
time (see e.g.~\cite{Berti:2006ew,Sesana:2007sh,Volonteri:2007fz}). To bracket
these uncertainties we focused on four representative models.

As a representative model with ``light'' black-hole seeds we considered the
Volonteri-Haardt-Madau (VHM) scenario~\cite{Volonteri:2002vz}, where light
seed black holes of $m_{\rm seed}\sim$ few hundred $M_\odot$ are produced as
remnants of metal-free stars at redshift $z\gtrsim 20$.  In an alternative
scenario, ``heavy'' seeds with $m_{\rm seed}\sim 10^5~M_\odot$ are formed as
the end-product of dynamical instabilities arising in massive gaseous
protogalactic disks in the redshift range $10\lesssim z \lesssim 15$. To allow
for the possibility of heavy seeds, we considered a variant of this scenario
proposed by Begelman, Volonteri and Rees (\cite{Begelman:2006db}, henceforth
BVR).  Both models (light and heavy seeds) can reproduce the AGN optical
luminosity function in the redshift range $1\lesssim z\lesssim 6$, but they
result in different coalescence rates of MBHBs and hence in different GW
backgrounds~\cite{Sesana:2007sh}. 

To bracket uncertainties in the growth of black holes by accretion, in both
the ``light seed'' (VHM) and ``heavy seed'' (BVR) cases we considered two
different accretion models. We adopted either the standard ``prolonged
accretion'' scenario, where accretion of material with constant angular
momentum axis rapidly spins up the holes~\cite{Bardeen:1970,Thorne:1974ve}, or
a ``chaotic accretion'' scenario~\cite{King:2006uu}, where accretion always
proceeds via very small and short episodes, caused by fragmentation of the
accretion disc where it becomes self-gravitating. Since counter-rotating
material spins black holes down more efficiently than co-rotating material
spins them up, and it is quite unlikely for mergers to produce rapidly
spinning holes \cite{Hughes:2002ei}, the chaotic accretion scenario implies
that black-hole spins are typically rather small~\cite{Berti:2008af}.

For all four models (with heavy/light seeds and efficient/chaotic accretion),
the spin resulting from individual merger events was determined using a
semianalytical fitting formula based on numerical relativity simulations of
the merger process~\cite{Rezzolla:2007rz}. This formula provides the modulus
of the final spin $\hjf$ given the binary's mass ratio $q=M_1/M_2$ and the
initial spins $\hjvec{i}$ $(i=1,2)$:
$
\hjf(q\,,\hj{1}\,,\hj{2}\,,\cos\alpha\,,\cos\beta\,,\cos\gamma) 
$.
The three angles $\cos\alpha=\hjvec{1} \cdot \hjvec{2}$, $\cos\beta =\hjvec{1}
\cdot \hat{\Lvec}$ and $\cos\gamma=\hjvec{2} \cdot \hat{\Lvec}$, where a hat
denotes unit vectors, describe the orientation of the initial spins relative
to the binary's orbital angular momentum $\Lvec$ (see~\cite{Rezzolla:2007rz}
for details).
The merger-tree simulations do not provide information on the source position
in the sky and on the orientation of the binary's angular momentum. However,
by averaging over different merger trees we can reasonably assume that all
angles are isotropically distributed in the sky. 

We further assume that spin alignment is not efficient, so that the spin
directions at merger are isotropically distributed. This assumption may be
violated if mergers commonly occur in gas-rich environments, and if the torque
exerted by the gas is efficient in producing alignment of the black hole's
angular momenta, as suggested in~\cite{Bogdanovic:2007hp}
(see~\cite{Berti:2008af} for more details).

\subsection{Details of the merger-tree implementation}

A merger tree traces the merger history that leads to a $z=0$ dark-matter halo
in a hierarchical cosmology. For each formation scenario, our sample is based
on $N_{\rm tree}\sim 10$ different halo masses~\cite{Lacey:1993iv} that span
the masses of interest (from a dwarf galaxy to a cluster of galaxies).  Each
merger tree is characterized by the mass of the parent halo at $z=0$ and its
Press-Schechter weight $W_{\rm PS}^{(k)}$ $(k=1,\dots,N_{\rm tree})$
\cite{Press:1973iz,Sheth:2001dp}, which is used to scale the results to the
(comoving) number density of sources. Furthermore, for each halo mass a
different number $N_{\rm real}^{(k)}$ of realizations of its merger history
have been performed, to take into account cosmic variance.  Typically,
large-mass halos have a smaller Press-Schechter weight (inherent in the
adopted cosmological model) and a smaller number of realizations (due to
computational burden).

For each tree $k$ we produced data files listing black-hole masses, spins and
redshifts corresponding to ``branches'' of the tree where a merger event
occurs. All quantities in these files are measured in the source frame, at
variance with the convention used in the MLDC (recall that $M=(1+z)M_{\rm
  source}$).  Including all merger trees and all realizations of each merger
tree, in a typical model such as VHM we have at least $\sim 5\times 10^4$
merging events, but many of these events will {\it not} be detectable by LISA.
Once we choose a criterion to select detectable binaries, e.g. by requiring
the SNR to be larger than some threshold value $\rho>\rho_{\rm th}$ in a
one-year observation time, the number of events at a given redshift $z$ per
comoving volume is
\be
N_{\rm com}(z)=
\sum_{k=1}^{N_{\rm tree}}
\sum_{j=1}^{N_{\rm mergers}^{(k)}}
\f{H(\rho^{(j,k)}(z)-\rho_{\rm th})\times W_{\rm PS}^{(k)}}{N_{\rm real}^{(k)}}\,,
\ee
where $H(x)$ is the Heaviside step function, and in practice we choose
$\rho_{\rm th}=10$. The number of observable events at $z=0$ per unit time
and redshift is then given by
\be\label{dNdzdt}
\f{d^2N}{dzdt}=
4\pi c N_{\rm com}(z)\left[\f{D_L(z)}{(1+z)}\right]^2\,,
\ee
where $D_L$ is the luminosity distance \cite{Haehnelt:1994wt}.

\section{Code Comparisons for Massive Black Holes with Higher-Harmonic
  Corrections\label{comparison}}

The importance to GW astronomy of higher-harmonic corrections to
post-Newtonian (PN) waveforms for massive black-hole inspirals has been
recognized long ago \cite{Sintes:1999cg,Hellings:2002si}, but recently there
have been several papers trying to quantify this effect more
precisely~\cite{Arun:2007qv,Arun:2007hu,Trias:2007fp,Porter:2008kn,Trias:2008pu}. 
Though the qualitative conclusions of all these papers were similar, the
quantitative estimates were different.  This has to do with differences in the
nature of waveforms used in the analysis (time versus frequency domain), the
different noise power spectral densities (PSDs) employed and different ways of
truncating the PN waveforms.  The main goal of the exercise presented in this
section is to bring together the results of various groups and compare them
using common values for the masses, the duration of the signal in the LISA
band, the noise PSD and the lower frequency cut-off of the detector.

To this end it was decided that each group would analyze four different
non-spinning MBHB sources.  The sources were described by the following
parameters: individual masses $(m_1, m_2)$, co-latitude and longitude
$(\theta_S,\phi_S)$, the polarization and inclination angles $(\psi,\iota)$ or
the orbital angular momentum variables $(\theta_L, \phi_L)$, the luminosity
distance $D_L$, the orbital phase at coalescence $\varphi_c$ and the instant
of coalescence $t_c$.  The given parameter sets were $(m_1/M_{\odot},
m_2/M_{\odot}, \theta_S, \phi_S, \psi, \iota, \theta_L, \phi_L,\varphi_c,
D_L/{\rm Gpc})=(3\times10^6, 10^6,
0.39845,5.10773,2.41199,2.77508,2.20726,2.85098,4.04657,25.8416)$ for case 1,
and $(3\times10^5, 10^5,1.62168,
0.920401,1.608812,0.949798,2.570387,0.977606,$ $2.697954,25.8416)$ for case 2.
Note that all angular variables are given in radians, and that the chosen
luminosity distance corresponds to a redshift $z=3$.  It was decided that each
group would use waveforms which were correct to 2PN order in both amplitude
and phase, and for each source we assumed one year of data. The time of
coalescence is $t_c=1$ year in cases 1a and 2a; in cases 1b and 2b, $t_c=1.05$
years, so that the merger occurs after the end of the data set.  We assumed
common MLDC~\cite{Arnaud:2006gm} conventions for the angles, and we adopted a
common list of physical constants.

More importantly, all groups agreed to use the noise curve that was used in
the second round of the MLDC \cite{Babak:2007zd}.
There are two components to the noise model: instrumental noise ($S_n$) and
confusion noise from the galactic foreground ($S_{\rm conf}$).  For our
comparison exercise, the (sky-averaged) instrumental noise is

\begin{equation}\label{noise-comparisons}
S_n(f) = \frac{1}{L^2}  
\left\{ 
\left[ 1 + \f{1}{2} \left(\frac{f}{f_*}\right)^2\right] S_p + 
\left[1+\left(\f{10^{-4}}{f}\right)^2\right]
\frac{4 S_a}{(2 \pi f)^4} 
 \right\} \,,
\end{equation}
where $f$ is in Hz, $L=5\times 10^9$~m is the armlength, $S_p = 4\times
10^{-22}$ m$^2$ Hz$^{-1}$ is the (white) position noise level, $S_a = 9\times
10^{-30}$ m$^2$~s$^{-4}$ Hz$^{-1}$ is the white acceleration noise level
(assumed equal to the red acceleration noise level) and $f_*=c/(2\pi L)$ is
the LISA arm transfer frequency. Note that the term $(f/f_*)^2/2$ in
Eq.~(\ref{noise-comparisons}) is not strictly due to an increase in shot noise
at high-frequency. Rather LISA's response to GWs falls rapidly for $f>f_*$,
and in the low-frequency approximation adopted in our parameter estimation
codes, this effect is typically accounted for by adjusting the noise curve in
this fashion.

The galactic confusion noise is estimated by direct simulation. A Nelemans
{\it et al.} \cite{Nelemans:2001hp,Nelemans:2003ha} population synthesis code
was used to produce a catalog of galactic binaries with periods shorter than
$2\times 10^4$ seconds. 
The catalog contained $2.6\times 10^7$ detached binaries and
$3.4\times 10^6$ interacting binaries. The LISA response to this foreground
was computed \cite{Cornish:2007if} and added to simulated instrument noise. A
Bayesian model selection technique was then used to determine which systems
could be individually identified and regressed from the data
\cite{Cornish:2007if}. The residual was then smoothed using a Savitzky-Golay
smoothing filter to produce an estimate for the galactic confusion noise. 
The confusion noise used in the comparisons is given by a piecewise fit to the
residual for a two-year nominal lifetime without interacting binaries:

\begin{equation}\label{confusion}
S_{\rm conf}(f) = 
\left\{ \begin{array}{ll}
10^{-44.62}f^{-2/3} & (f\leq 10^{-3})\,, \\
10^{-50.92}f^{-4.4} & (10^{-3}<f<10^{-2.7})\,,  \\
10^{-62.8}f^{-8.8}  & (10^{-2.7}<f<10^{-2.4})\,,  \\
10^{-89.68}f^{-20}  & (10^{-2.4}<f<10^{-2})\,,  \\
0                 & (f\geq 10^{-2})\,,
\end{array} \right.
\end{equation}
where $f$ is in Hz. While there were differences between the way waveforms
were treated, the main difference between the various groups was the
analytical noise curve that was used.  Once a common noise curve was agreed
upon, most of the disagreement between groups disappeared.  A final tidying up
of the slight disagreements between waveform models produced a concrete
verification of the results.

In the next few sections, we outline the different waveform approximations
that were used.  Two groups used time-domain waveforms, and two groups used
frequency-domain waveforms. To begin with, it seemed reasonable to make
comparisons between either time-domain or frequency-domain based codes.  We
then compared between the time- and frequency-domain results.  The main
differences between waveform models are summarized in Table~\ref{tab:compar}.

\begin{table}[htb]
  \caption{\label{tab:compar}Comparisons of the different methods employed in
    the analysis.  All waveforms used a 2 PN phase approximation for the
    study.}
\begin{tabular}{@{}lllll}
\br
Group         & \multicolumn{1}{c}{AISSV} & \multicolumn{1}{c}{MM} & \multicolumn{1}{c}{CP} & \multicolumn{1}{c}{TS}  \\
\mr
Domain : & Frequency & Time & Time & Frequency   \\
Amplitude: & Full PN expansion& MLDC taper, & MLDC taper. & Full PN expansion.   \\
           & and truncation. & Hann window. &            & No truncation.\\
Spin : & No & Yes & No & Yes\\
Spin Precession : & No & Yes & No & No\\
\br
\end{tabular}
\end{table}

\subsection{Time-Domain Waveforms}

Two independent time-domain codes were used for the analysis.  The first was
used by Cornish and Porter (CP)~\cite{Porter:2008kn} and is based on a 2PN
waveform in both phase and amplitude, derived by Blanchet, Iyer, Will and
Wiseman~\cite{Blanchet:1996pi}.  This code uses analytic expressions for the
evolution of both the orbital phase and frequency.  The code developed by the
Montana/MIT group (MM)~\cite{ltools} is based on comparable-mass Kerr inspiral
waveforms as developed by Apostolatos, Cutler, Sussman and
Thorne~\cite{Apostolatos:1994mx}.  This code used coupled ordinary
differential equations to evolve the spin and spin-precession equations.  For
this study, the spin is set to zero.  In order to dampen ringing in the
Fourier domain, both waveforms use the standard MLDC hyperbolic truncation
function~\cite{Arnaud:2006gm}.  This function begins to truncate the waveform
at an orbital separation of $7M$. The MM code also uses a Hann window to
prevent leaking of higher harmonics into each other in the Fourier transform.

\begin{table}[htb]
  \caption{\label{tab:cpch}Fractional differences in the claimed
    parameter-estimation accuracy between the CP and MM models. We list the
    SNR and the accuracy in (from left to right): individual masses, time to
    coalescence, luminosity distance, sky angles, orbital inclination,
    polarization angle and orbital phase at coalescence.   Note that for each
    quantity $\Delta \lambda$, the value in the Table should be multiplied by
    $10^{-4}$.}
\lineup
\begin{tabular}{@{}lllllllllll}
\br
Case         & \multicolumn{1}{c}{$\Delta ({\rm SNR})$} &
\multicolumn{1}{c}{$\Delta \ln( m_{1})$} & \multicolumn{1}{c}{$\Delta
\ln(m_{2})$} & \multicolumn{1}{c}{$\Delta \ln(t_c)$} &
\multicolumn{1}{c}{$\Delta \ln(D_L)$ } & \multicolumn{1}{c}{$\Delta
\cos\theta_S$} & \multicolumn{1}{c}{$\Delta \phi_S$ } &
\multicolumn{1}{c}{$\Delta \iota$} & \multicolumn{1}{c}{$\Delta \psi$ } &
\multicolumn{1}{c}{$\Delta \varphi_c$} \\
\mr
1a & 1.25 & 5.61 & 5.84 & 6.27 & 175.07 & 14.57 & 30.09 & 248.35 & 125.69
& 57.29  \\
1b & 1.73 & 2.90 & 2.89 & 3.55 & 32.75 & 0.48 & 3.61 & 41.29 & 37.28 &
3.18  \\
2a & 18.49 & 19.99 & 20.05 & 74.95 & 566.11 &365.62 & 90.57 & 800.23 &
385.82 & 146.92  \\
2b & 0.77 & 0.97 & 0.98 & 0.58 & 3.25 & 1.04 & 1.08 & 1.95 & 8.55 & 0.79  \\
\br
\end{tabular}
\end{table}

In Table~\ref{tab:cpch} we show the difference in the claimed
parameter-estimation accuracies between the two methods.  We can see that this
difference is typically of the order of a percent, and sometimes much less
than that: for example, in case 1a the fractional difference between codes in
calculated mass accuracy is $\sim 6\times 10^{-4}$.

\subsection{Frequency-Domain Waveforms}

The two groups using frequency domain codes were Arun, Iyer, Sathyaprakash,
Sinha and Van Den Broeck (AISSV)~\cite{Arun:2007qv, Arun:2007hu} and Trias and
Sintes (TS)~\cite{Trias:2007fp,Trias:2008pu}.  While both groups were using
the Stationary Phase Approximation (SPA)~\cite{Thorne} to obtain
frequency-domain waveforms, the treatment of some aspects of waveform
generation was different.  The SPA amplitude of the $k$th harmonic is composed
of a ratio of PN series.  The AISSV group expanded the ratio of these series
and truncated at the required 2 PN order.  They also used an analytic
expression to calculate the starting frequency of the waveform in the
detector. The TS group expanded both PN amplitude series as was necessary, but
did not truncate the ensuing ratio.  They also used a numerical inversion of
the PN series for $t(f)$ at $t = 0$ to find the starting frequency.  Their
code allowed for non-zero spins, but not for spin precession.  For code
comparisons, the spins were set to zero.

\begin{table}[hbt]
  \caption{\label{tab:aissvts}Fractional differences in the claimed parameter-estimation 
    accuracy between the AISSV and TS models.  The parameters we
    consider are time to coalescence, chirp mass, luminosity distance, sky and
    orbital angular resolution, and orbital phase at coalescence.  The angular
    resolution variables are defined by
    $\Delta\Omega_i=2\pi[\Sigma^{\theta\theta}\Sigma^{\phi\phi}-\left(\Sigma^{\theta\phi}\right)^{2}]^{1/2}$,
    where $\Sigma^{ij}$ correspond to the elements of the variance-covariance
    matrix, and $i = S,L$.  Note that again, values in the Table should be
    multiplied by $10^{-4}$.}
\begin{indented}\lineup
\item[]\begin{tabular}{@{}lllllll}
\br
Case         & \multicolumn{1}{c}{$\Delta t_c$} &
\multicolumn{1}{c}{$\Delta \ln( M_{c})$} & \multicolumn{1}{c}{$\Delta
\ln(D_{L})$} & \multicolumn{1}{c}{$\Delta\Omega_S$} &
\multicolumn{1}{c}{$\Delta\Omega_L$ } & \multicolumn{1}{c}{$\Delta
\varphi_c$} \\
\mr
1a & 3.87 & 16.40 & 8.35 & 1.04 & 0.46 & 13.18  \\
2a & 10.22 & 24.81 & 0.00 & 4.69 & 1.34 & 7.08  \\
\br
\end{tabular}
\end{indented}
\end{table}

In Table~\ref{tab:aissvts} we compare the two groups' results.  For this
exercise, the AISSV group agreed to use the same PN truncation of the
amplitude as the TS group. We can see that in all cases the results differ by
less than 1\%.

\subsection{Comparison Between Time- and Frequency-Domain Waveforms}

In initial comparisons between time- and frequency-domain codes we observed
discrepancies in the SNRs.  A visual comparison was made by inverse Fourier
transforming the frequency domain waveforms and comparing with the time-domain
waveforms.  It turned out that the definition of the phase at coalescence is
slightly different in the time and frequency domains.  In order to match
waveforms, a slight correction of the phase at coalescence was needed in the
frequency-domain waveforms.

Another aspect affecting the results was the fact that both time-domain models
used a truncation function to smoothly reduce the signal to zero at the end of
the inspiral phase.  As this truncation function was also affecting the
comparison between the time- and frequency-domain approaches, it was decided
that an upper frequency cutoff of 0.1 mHz would be applied to case 1a, and an
upper frequency cutoff of 1 mHz would be applied to case 2a.  This nullified
the effect of the time-domain taper.  In Table~\ref{tab:all2} we compare the
four sets of results from the MM, AISSV, modified AISSV and TS entries.  We
can see that there is excellent agreement between all groups.

\begin{table}[htb]
  \caption{\label{tab:all2}Parameter-estimation accuracies for Case 2a for
    both time- and frequency-domain codes, assuming an upper frequency cutoff
    of $10^{-3}$ Hz. We present results for the MM, TS and AISSV groups.  The
    AISSV* entry is the code modified to match the TS formalism.}
\begin{indented}\lineup
\item[]\begin{tabular}{@{}lllll}
\br
Group          & \multicolumn{1}{c}{MM} & \multicolumn{1}{c}{TS} & \multicolumn{1}{c}{AISSV} & \multicolumn{1}{c}{AISSV*} \\
\mr
${\rm SNR}$ 	& \multicolumn{1}{c}{101.2}	&	\multicolumn{1}{c}{100.9}	&	\multicolumn{1}{c}{99.41}	&	\multicolumn{1}{c}{100.8}	\\
$\Delta \ln(M_c)$ & \m $4.651\times10^{-5}$   & \m   $4.671 \times10^{-5}$  & \m $4.678 \times10^{-5}$ & \m $4.659 \times10^{-5}$\\
$\Delta t_c ({\rm s})$     & \m28.17  & \m27.14 & \m27.41 & \m27.15\\
$\Delta \ln(D_L)$     & \m0.0673                    & \m0.0674         & \m0.0677 & \m0.0675\\
$\Delta \Omega_S ({\rm deg}^2)$  & \m $17.55$    & \m   $17.56$  & \m $17.53$ & \m $17.52$\\
$\Delta \Omega_L ({\rm deg}^2)$ & \m $17.05$    & \m   $17.13$  & \m $17.57$ & \m $17.29$\\
$\Delta \varphi_c$    &          \m0.1225          &  \m0.1211    & \m0.1213  & \m0.1202\\
\br
\end{tabular}
\end{indented}
\end{table}

Finally, only the MM time-domain model had the ability to include at the same
time higher harmonics, spins and spin precession.  While no comparison was
possible with the other groups taking part in this exercise, the MM code has
been tested by comparison with an MLDC code written by Babak using the Kidder
model to evolve spins~\cite{ltools,Kidder:1995zr}.  Because only the MM model
had a working Fisher matrix, the comparison was done by generating waveforms
for both models and then subtracting them from each other.  This test yielded
very small residuals and confirmed the validity of the MM treatment of spins
and spin precession.

\section{Results\label{results}}

In the foregoing sections we discussed the merger-tree models used in this
work (Sec.~\ref{merger}) and described the cross-validation of the different
parameter-estimation codes (Sec.~\ref{comparison}).  Here we will present the
source distributions resulting from our merger-tree models, and use them for
Monte Carlo simulations of LISA observations of MBHBs. One purpose of this
exercise was to evaluate the impact of different design choices on LISA's
ability to accurately measure and localize MBHBs. Therefore, for each source
distribution we computed parameter accuracies using two different LISA noise
curves: the ``baseline'' and ``6-link'' noise models, described below.

\subsection{Source distributions}\label{sourcedistributions}

To assess what science LISA can do with MBHB mergers it is essential to
understand the distribution of sources as a function of redshift, masses and
spins of the component black holes. We evaluated parameter accuracies for the
four different merger models described in Sec.~\ref{merger}. We shall call the
models SE (small seeds, efficient accretion), SC (small seeds, chaotic
accretion), LE (large seeds, efficient accretion) and LC (large seeds, chaotic
accretion).

\begin{figure}
\includegraphics[width=2in,angle=-90]{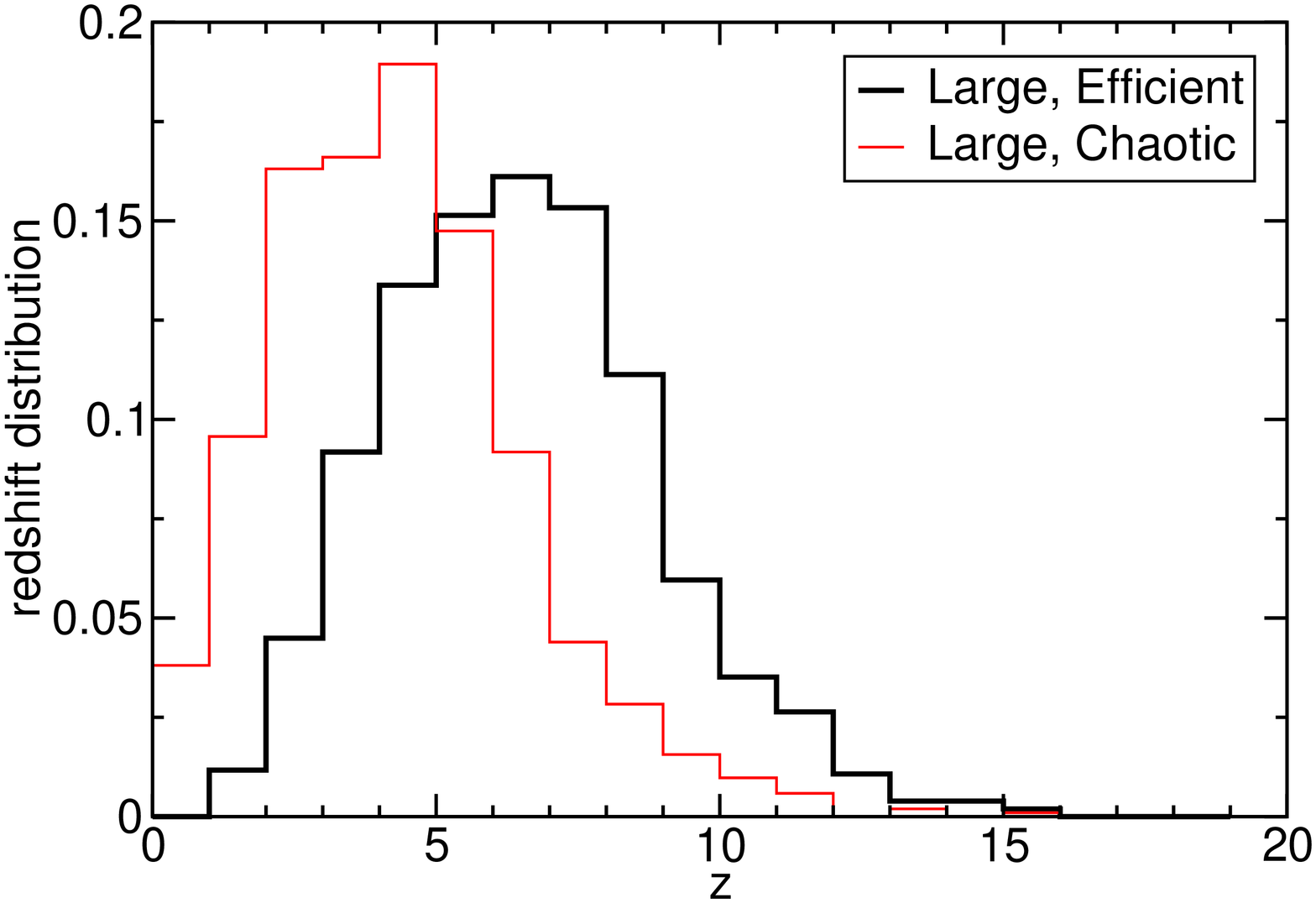}
\includegraphics[width=2in,angle=-90]{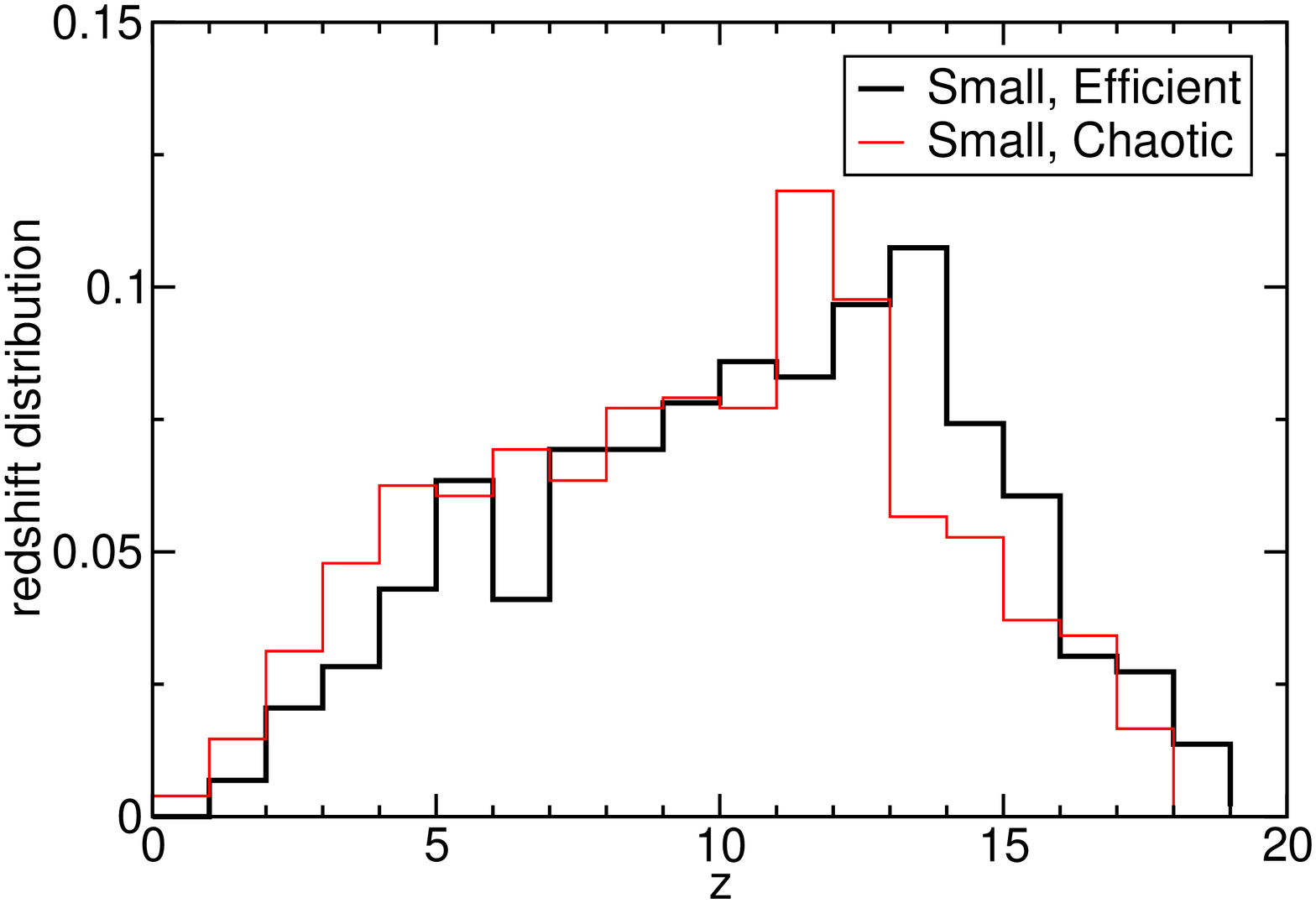}
\includegraphics[width=2in,angle=-90]{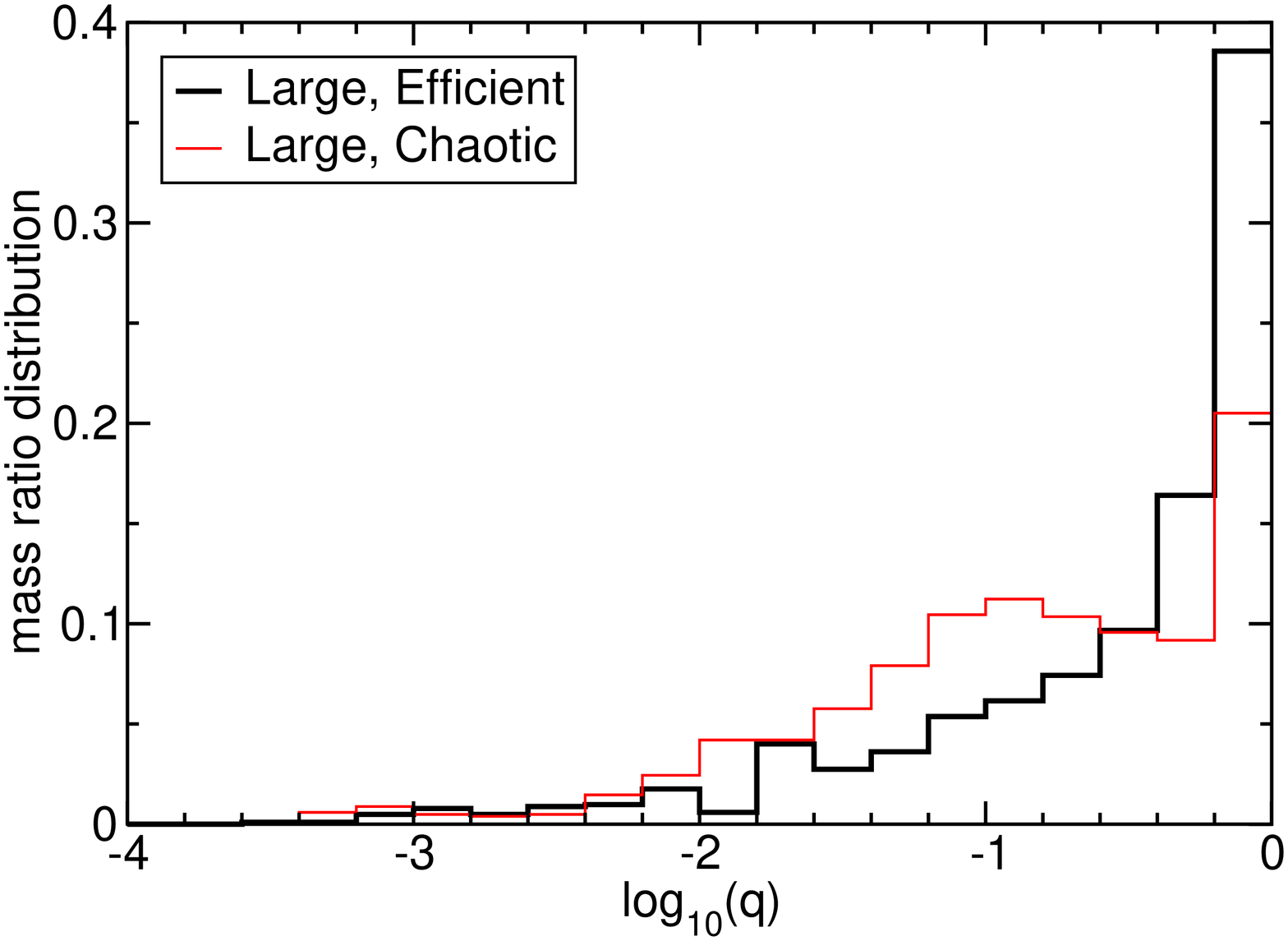}
\includegraphics[width=2in,angle=-90]{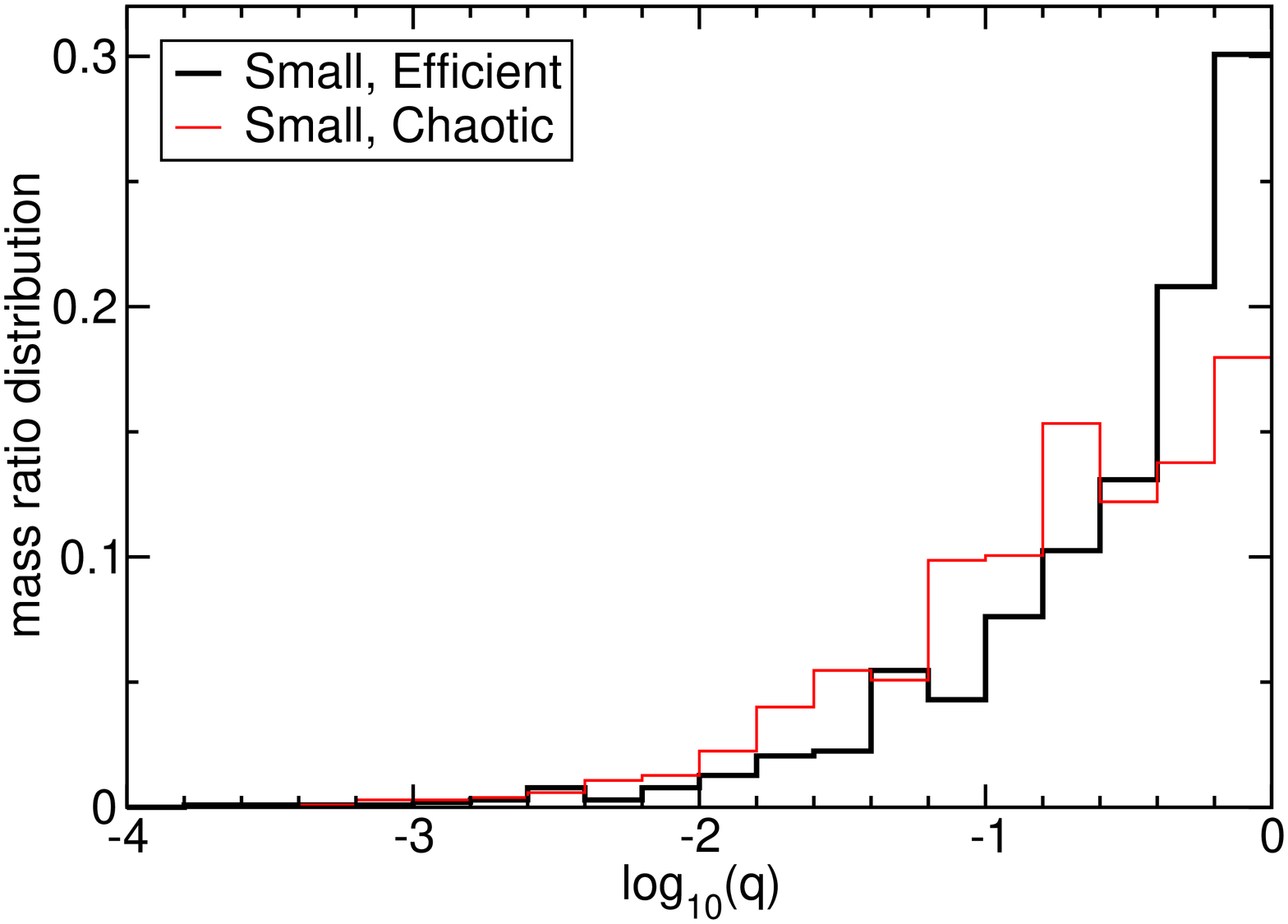}
\caption{The distribution of MBHB mergers as a function of redshift $z$ (top
  panels) and (logarithmic) mass ratio $\log(q)$ (bottom panels) for
  large-seed models (left) and small-seed models (right). Thick (black)
  histograms refer to efficient accretion, thin (red) histograms to chaotic
  accretion.}
\label{fig:z-qDistr}
\end{figure}

For large-seed models most mergers occur in the redshift range $3\lsim z \lsim
8$, with a peak around $z\sim 5$ (Fig.\ \ref{fig:z-qDistr}, top left
panel). In the case of small seeds, mergers are roughly uniform in $z$ over
the range $4 \lsim z \lsim 15$, with a peak around $z\sim 12$ (Fig.\
\ref{fig:z-qDistr}, top right panel).  Large-seed models are likely to produce
more symmetric binaries (which produce larger SNR for fixed total mass). In
contrast, small-seed models lead to more asymmetric systems (see
Fig.~\ref{fig:z-qDistr}, bottom panels).  Thus, although small-seed MBHB
mergers could occur more frequently, a smaller fraction would be observed by
LISA due to their smaller total mass and less symmetric mass ratios.

Spin precession usually improves parameter-estimation accuracies
\cite{Vecchio:2003tn,Lang:1900bz}. LE models are the only ones that produce
binaries in which both black holes generally have large spins. In SE models
the spin is large for the more massive black hole, but it is often negligible
for the smaller hole. In chaotic accretion models (LC and SC) spins are always
negligible. For a more detailed discussion of spin evolution and observational
implications, see \cite{Berti:2008af}.

\subsection{Baseline and 6-link noise models}

\begin{figure}[htb]
\begin{tabular}{c}
\includegraphics[width=2in,angle=-90]{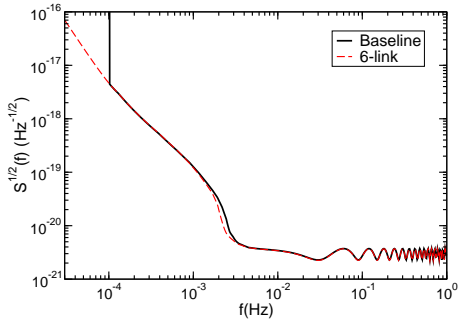}
\end{tabular}
\caption{Baseline and 6-link noise models (including both instrumental and
  confusion noise) used in the merger-tree analysis.}
\label{fig:noise}
\end{figure}

To evaluate the impact of different LISA design choices on the mission’s
science performance, for each source distribution we computed parameter
accuracies using two different LISA configurations and associated noise
curves: a ``6-link'' configuration, which allows the construction of all three
independent TDI channels, and a ``baseline'' configuration of 4 links,
producing a single Michelson channel. The instrumental noise $\tilde S_n$ is
similar for the baseline and 6-link configurations, save for the location of
the low frequency ``wall'', which is at $10^{-4}$ Hz for the baseline and
$3\times 10^{-5}$~Hz for the 6-link model.
The instrumental noise in the Michelson A and E channels is given by
\begin{equation}\label{noise}
\tilde S_n(f) = \frac{1}{12 L^2}  
\left\{ ( 2 + k) \tilde S_p + \frac{1 + k + k^2}{(2 \pi f)^4} 
S_a \left[1+\left(\frac{10^{-4}}{f}\right) \right] \right\} \,,
\end{equation}
where $k = \cos(f/f_*)$. Here $\tilde S_p = 3.24\times
10^{-22}$~m$^2$~Hz$^{-1}$ is slightly different from the value adopted in
Eq.~(\ref{noise-comparisons}), corresponding to the baseline currently adopted
by the LIST, and the second term in square parentheses is the pink
acceleration noise level. The confusion noise estimates $\tilde S_{\rm
  conf}(f)$ used in the analysis of MBHB merger trees, unlike those for the
comparison exercise described in Sec.~\ref{comparison}, did include
interacting binaries. The confusion noise for the baseline is slightly higher
than for the 6-link configurations as the latter is a factor of $\sim 2$ more
sensitive, and has a five- versus three-year nominal lifetime. The total
(instrumental plus confusion) noise spectral densities for these models are
plotted in Figure~\ref{fig:noise}.

\subsection{Parameter accuracies}

LISA's parameter-estimation capabilities were calculated using the MM code to
compute the covariance matrix. As discussed in Sec.~\ref{comparison}, we have
cross-validated their code against other codes finding excellent
agreement. The advantage of the MM code is that it incorporates not only
high-order amplitude corrections, but also precessional effects due to spin.
However, even the MM code does not take advantage of the merger and ringdown
parts of the signal. Therefore, the parameter accuracies obtained in this work
probably {\em underestimate} LISA's potential performance.

The statistics of detectability and parameter accuracies are based on a Monte
Carlo simulation in which we randomly selected 1024 events with the
distributions discussed in Sec.~\ref{sourcedistributions}. We eliminated from
the sample all binaries with total (locally-measured) mass larger than $3
\times 10^7 M_{\odot}$, since a) the code often crashed for such high masses,
b) these cases represented a very small fraction of the population, and c)
their inspirals take place almost entirely outside the LISA band, and so would
be very poorly localized in any case. We then calculated 2048 covariance
matrices (1024 for each of the two LISA noise models) on JPL's Cosmos
supercomputer. About two percent of the jobs did not finish within a preset 3
hour period, and some 0.1\% of the simulations had failures; such cases were
simply discarded\footnote {Our covariance matrices are computed in 15
  dimensions. Because of large covariances between parameters (often close to
  $\pm 1$) the Fisher information matrix is ill-conditioned, and we believe
  our occasional failures to arise when numerical inversion of the Fisher
  matrix fails within the available numerical precision. We will explore the
  cause in more detail in the future.}.

\begin{table}[htb]
  \caption{For each merger-tree model (SE, SC, LE and LC) we list: the total
    number $N$ of MBHB events in LISA's past light cone in a one-year
    observation; the number of events $N_{\rm det}$ detectable with SNR larger
    than 10 in one year; the number for which the error in the luminosity
    distance $D_L$ is 10\% or less; the number that is localizable within 1
    and 10 deg$^2$ ($N_{\rm 1\, deg^2}$ and $N_{\rm 10\, deg^2}$,
    respectively); the number that can be resolved to within 10 deg$^2$ with
    less than 10\% errors in $D_L$ ($N_{{\rm 10\,deg^2},\,10\%D_L}$), and the
    same for 1 deg$^2$ in angular resolution and 1 \% error in $D_L$ ($N_{{\rm
        1\,deg^2},\,1\%D_L}$). Results for the 6-link model are followed by
    those for the baseline model (within parentheses).}
\label{tab:PE}
\begin{center}
\begin{tabular}{cccccccc}
\hline
\hline
Model  & $N$  & $N_{\rm det}$ & $N_{10\% D_L}$ & $N_{\rm 10\,deg^2}$ & $N_{{\rm 10\,deg^2},\, 10\%D_L}$ & $N_{\rm 1\, deg^2}$ & $N_{{\rm 1\, deg^2},\, 1\%D_L}$ \\ 
\hline
SE     & 80   &  33 (25)          & 21 (8.0)       & 8.2 (1.5)        & 7.9 (1.1)      & 2.2 (0.6)  & 1.7 (0.1)  \\
SC     & 75   &  34 (27)          & 17 (4.4)       & 6.1 (0.4)        & 5.5 (0.4)      & 1.3 (0.1) & 1.3 (0.1)  \\
\hline
LE     & 24   &  23 (22)          & 21 (7.7)       & 10 (0.8)         & 10 (0.7)       & 2.2 (0.1)  & 1.2 (0.05)  \\
LC     & 22   &  21 (19)          & 14 (4.3)       &  6.5 (0.5)       &  5.4 (0.5)     & 1.8 (0.04)  & 1.0 (0.1)  \\
\hline
\hline
\end{tabular}
\end{center}
\end{table}

For our four merger-tree models (as listed in column 1 of Table \ref{tab:PE})
we first give the total number $N$ of mergers in LISA's past light-cone during
a one-year observation (column 2) and the number of events $N_{\rm det}$ that
are detectable in one year with SNR larger than 10 (column 3).  The rest of
the columns show the number of observable events with: error in $D_L$ of 10\%
or less (column 4); angular resolution $\Delta \Omega_S<10$ deg$^2$ (column
5); $\Delta \Omega_S<10$ deg$^2$ and error in $D_L$ of 10\% or less (column
6); $\Delta \Omega_S<1$ deg$^2$ (column 7); $\Delta \Omega_S<1$ deg$^2$ and
error in $D_L$ of 1\% or less (column 8).

LISA will detect quite a good fraction of the mergers which occur in the
universe: almost all mergers will be detected in large-seed scenarios, and
nearly half of all mergers in small-seed scenarios. This is because large seed
black holes lead to more massive MBHBs with mass ratio $q$ close to one, so
they can be seen out to larger redshifts. Since the number of mergers in
LISA's past light cone is three times larger in the case of small seeds, the
number of detectable events will be similar in both cases. Table \ref{tab:PE}
shows that the number of detectable events (about $30$ per year in small-seed
models, and about $20$ per year in large-seed models) seems to be quite
insensitive to the details of accretion and spin evolution.  The number of
detectable events is slightly smaller for the baseline model than for the
6-link model.

Measurements of $D_L$ accurate within 1\% would allow us to determine the dark
energy equation-of-state parameter $w$ to better than 1\%, but this would not
be possible if the accuracy in $D_L$ were of order 10\% \cite{Arun:2007hu}.
The typical error in $D_L$ due to weak lensing is in the range 5--10\% at $z
\sim 2$ \cite{Holz:2005df,Kocsis:2005vv}. Ideal MBHB sources to do precision
cosmology (``standard sirens'') would be in the redshift range (say)
$0.5<z<0.8$, where lensing errors would presumably be of order
2--3\%. Therefore, an accuracy in $D_L$ of about 1\% is needed if we want our
standard sirens to be limited only by lensing, and not by the random errors
due to LISA's instrumental noise. The temporal coincidence with any prompt
electromagnetic counterpart may offer the best chance of identifying a unique
host galaxy, especially if we can locate the source within one (or a few)
square degrees \cite{Kocsis:2007yu}.

Table \ref{tab:PE} clearly shows that with the merger-tree models and
gravitational waveforms used in this study, both the baseline design and the
6-link model have the potential for precision cosmology. However, the 6-link
configuration performs significantly better than the baseline design.
With the 6-link configuration LISA can get the luminosity distance to within
10\% for the majority of detectable events. This is not really surprising,
since the luminosity distance does not have a strong correlation with the
phasing parameters and our detection threshold is $\rho_{\rm th}=10$.
However, only a fourth to half of all events can be resolved within a 10
deg$^2$ error box. Once an event has a good sky resolution it will also have a
good accuracy in $D_L$, as evidenced by the similarity of the numbers in
columns 5 and 6: this is due to the high correlation between luminosity
distance and angular parameters of the source (see
e.g.~\cite{Berti:2004bd,Lang:1900bz}).

The fraction of events that can be localized to within 1 deg$^2$ is far
smaller. A few events can be localized to such accuracy using one-year
observations and the 6-link noise curve, while the statistics are too low to
determine if events could be localized to such accuracy with the baseline
design.
Finally, while we have considered only 6-link and 4-link LISA configurations,
it is possible that LISA will operate with 5 links for some substantial time.
The important point is that for MBHB mergers, LISA’s sensitivity and parameter
extraction accuracy with 5 links should be nearly as good as with 6
\cite{Vallisneri:2007xa}. Therefore it is quite plausible that $\sim 1$ event
per year could have angular resolution and accuracy in luminosity distance
good enough to measure the dark-energy equation-of-state parameter $w$
\cite{Arun:2007hu}.

\section{Conclusions\label{conclusions}}

We compared four largely independent codes for calculating LISA's 
parameter-estimation capabilities. All codes are based on the Fisher-matrix
approximation, but in the past they used somewhat different signal models,
source parametrizations and noise curves. We showed that once these
differences are removed, the four codes give results in extremely close
agreement with each other. 

Using a code that includes both spin precession and higher harmonics in the GW
signal, we presented a preliminary exploration of LISA's ability to do
precision cosmology using four different merger-tree models and two different
sensitivity curves.  For the merger history of galaxies we used either
``small'' seed black holes ($m_{\rm seed} \sim 100 M_\odot$) or ``large''
seeds ($m_{\rm seed}\sim 10^5M_\odot$).  In each case we adopted two different
accretion scenarios (efficient or chaotic accretion), thereby giving us four
models for the birth, growth and spin evolution of massive black holes.  We
then employed a carefully tested tool to compute the parameter accuracies for
events that LISA can detect with an (amplitude) SNR of at least 10.

Our study shows that LISA has good potential for carrying out precision
cosmology.  There will be about 20 sources with a modest distance measurement
(to within 10 \%) and about 10 sources with a modest sky resolution (10
deg$^2$). Even more interestingly, each year LISA may observe a few sources
with excellent sky resolution (1 deg$^2$) and luminosity distance measurements
(to within 1\%). Such accuracies are good enough to measure the dark-energy
equation-of-state parameter $w$ to within a percent, the dominant source of
uncertainty being errors in the luminosity distance from weak lensing of GW
events \cite{Holz:2005df,Kocsis:2005vv}.

The waveforms used in our study included important physical effects, such as
high-order PN corrections in amplitude and phase and precessing spins. However
our waveforms were incomplete in one respect: they did not include the merger
and ringdown parts of the signal. Future work will include ringdown either in
a phenomenological way (by estimating the final black hole's parameters from
the inspiral parameters, and then using the ringdown measurement formalism
developed in \cite{Berti:2005ys}) or by ``stitching'' PN inspiral waveforms to
a set of numerical relativity waveforms. As shown recently by Babak {\it et
  al.}  \cite{Babak:2008bu} (see also \cite{Thorpe:2008wh}) the inclusion of
the merger and ringdown signal is expected to improve parameter estimation in
two ways: by enhancing the signal-to-noise ratio and by increasing the signal
bandwidth. Future studies will need to explore the contribution of merger and
ringdown signals to detection rates and parameter-estimation accuracy. They
should also explore a broader class of merger-tree models to get a more
realistic evaluation of LISA's ability to do precision cosmology.

The LISAPE Taskforce is presently developing tools similar to those described
in this paper to explore measurement accuracies from extreme mass-ratio
inspirals (EMRIs) of compact objects into massive black holes. This work will
be reported elsewhere.

\section{Acknowledgments}

We wish to thank Michele Vallisneri for very helpful interactions.
E.B.'s and C.C.'s work was carried out at the Jet Propulsion Laboratory (JPL),
California Institute of Technology, under contract with the National
Aeronautics and Space Administration.  C.C.'s work was partly supported by
NASA Grant NNX07AM80G. E.B.'s research was supported by an appointment to the
NASA Postdoctoral Program at JPL, administered by Oak Ridge Associated
Universities through a contract with NASA.
N.C. is supported by NASA grant NNX07AJ61G.
A.S. and M.T. would like to thank the support of the Max-Planck Society, the
Spanish Ministerio de Educaci\'{o}n y Ciencia Research Projects
FPA-2007-60220, HA2007-0042, CSD207-00042 and the Govern de les Illes Balears,
Conselleria d'Economia, Hisenda i Innovaci\'{o}.
S.A.H and R.N.L. have been supported by NASA Grants NNG05G105G and NNX08AL42G,
as well as NASA Contract No.\ 1291617 and the MIT Class of 1956 Career
Development Fund.
I.M. was partially supported by NASA ATP Grant NNX07AH22G to Northwestern
University.
E.K.P. would like to thank the DLR (Deutsches Zentrum f\"ur Luft- und
Raumfart) for support during this work.
Research at Cardiff was supported in part by PPARC grant PP/F001096/1.
The Monte Carlo parameter-estimation results presented here were generated on
JPL's Cosmos supercomputer.
\copyright 2008. All rights reserved.

\end{document}